\documentclass[12pt]{iopart}
\usepackage{iopams}
\usepackage{graphicx}
\usepackage{bm}
\usepackage{dcolumn}
\usepackage{color}
\usepackage[latin9]{inputenc}
\usepackage{url}
\usepackage{subfigure}
\usepackage{float}
\usepackage{longtable}
\usepackage{epstopdf}

\sloppypar

\newcommand{\raiseentry}[1]{\smash{\raise 0.7 em \hbox{#1}}}

\newenvironment{equationarray*}
{\arraycolsep 0.14 em
\begin{eqnarray*}}
{\end{eqnarray*}}

\begin{document}

\title[Classification methods for noise transients in advanced gravitational-wave detectors]{Classification methods for noise transients in advanced gravitational-wave detectors II: performance tests on Advanced LIGO data.}

\author{ Jade Powell$^1$, Alejandro Torres-Forn\'e$^2$, Ryan Lynch$^3$, Daniele Trifir\`o$^4$ $^5$, Elena Cuoco$^6$ $^7$, Marco Cavagli\`a$^5$, Ik Siong Heng$^1$ and Jos\'e A. Font$^2$ $^8$}

\address{$^1$ SUPA, Institute for Gravitational Research, School of Physics and Astronomy, University of Glasgow, Glasgow, G12 8QQ, Scotland, United Kingdom.}
\address{$^2$ Departamento de Astronom\'{\i}a y Astrof\'{\i}sica, Universitat de Val\`encia, Dr. Moliner 50, 46100, Burjassot (Val\`encia), Spain.}
\address{$^3$ Massachusetts Institute of Technology, 185 Albany St, 02139 Cambridge USA}
\address{$^4$ Dipartimento di Fisica E. Fermi, Universit\`a di Pisa, Pisa 56127, Italy.}
\address{$^5$ Department of Physics and Astronomy, The University of Mississippi, University, MS 38677, USA.}
\address{$^6$ European Gravitational Observatory (EGO), Via E Amaldi, I-56021 Cascina, Italy}
\address{$^7$ Istituto Nazionale di Fisica Nucleare (INFN) Sez. Pisa Edificio C - Largo B. Pontecorvo 3, 56127 Pisa, Italy.}
\address{$^8$Observatori Astron\`omic, Universitat de Val\`encia, C/ Catedr\'atico Jos\'e Beltr\'an 2, 46980, Paterna (Val\`encia), Spain.}
\date{\today}


\begin{abstract}
The data taken by the advanced LIGO and Virgo gravitational-wave detectors contains short duration noise transients that limit the significance of astrophysical detections and reduce the duty cycle of the instruments. As the advanced detectors are reaching sensitivity levels that allow for multiple detections of astrophysical gravitational-wave sources it is crucial to achieve a fast and accurate characterization of non-astrophysical transient noise shortly after it occurs in the detectors. Previously we presented three methods for the classification of transient noise sources. They are Principal Component Analysis for Transients (PCAT), Principal Component LALInference Burst (PC-LIB) and Wavelet Detection Filter with Machine Learning (WDF-ML). In this study we carry out the first performance tests of these algorithms on gravitational-wave data from the Advanced LIGO detectors. We use the data taken between the 3rd of June 2015 and the 14th of June 2015 during the 7th engineering run (ER7), and outline the improvements made to increase the performance and lower the latency of the algorithms on real data. This work provides an important test for understanding the performance of these methods on real, non stationary data in preparation for the second advanced gravitational-wave detector observation run, planned for later this year. We show that all methods can classify transients in non stationary data with a high level of accuracy and show the benefits of using multiple classifiers.
\end{abstract}


\section{Introduction}
\label{section:introduction}

The advanced Laser Interferometer Gravitational-Wave Observatory (aLIGO) detectors are two 4km interferometers at Hanford, Washington (H1) and Livingston, Louisiana (L1) \cite{2010CQGra..27h4006H, 0264-9381-32-7-074001}. The Italian 3km interferometer Virgo is expected to join the advanced detector network early in 2017 \cite{2008CQGra..25k4045A}. The detector duty cycle and sensitivity to astrophysical signals will be determined by noise sources created by the instruments and the environment. In particular, as the detector noise is non-Gaussian short-duration transients will limit the sensitivity of searches for transient astrophysical sources such as compact binary coalescences \cite{Allen:2005cz}.

The first aLIGO observation run (O1) began autumn 2015. On the 14th September 2015 the aLIGO and Virgo teams detected gravitational waves from the binary black hole system GW150914 \cite{2016PhRvL.116f1102A}. A second binary black hole detection was made on the 26th of December \cite{PhysRevLett.116.241103}. An extensive study of the noise transients, which occurred in the data containing the detections, was carried out for the validation of the signals \cite{2016arXiv160203844T}. As the advanced detector network approaches its design sensitivity, the number of detections is expected to increase. Adding more detectors to the network increases the number of possible noise sources and the time it will take to identify their origin. Transients which occur in any one detector will limit the joint analysis time for the network. Understanding the sources of noise transients in the detectors will become increasingly more important with a latency of a few hours.  

The detectors contain many environmental and instrumental sensors, which produce auxiliary channels of data that can be used to monitor the detector behaviour and track the causes of short-duration noise artifacts. Auxiliary channels that are not sensitive to gravitational waves can be used to identify noise transients, also known as ``glitches", in the detector output and veto those events \cite{2011CQGra..28w5005S, 2014PhRvD..89l2001A, 2014arXiv1410.7764T}. Classification and categorization of transients using individual channels of data may provide valuable clues for the identification of their sources, which can aid in efforts to eliminate them \cite{2015CQGra..32u5012P, 1742-6596-243-1-012006}. So far classification has mainly been achieved by visual inspection of spectrograms of the transients, but automatic classification is essential for future detections of astrophysical gravitational-wave signals.

Three methods for fast classification of transients have been developed for the analysis of aLIGO and Virgo data. They are Principal Component Analysis for Transients (PCAT), Principal Component LALInference Burst (PC-LIB) and Wavelet Detection Filter with Machine Learning (WDF-ML). Previous work has shown that these methods can classify artificial data sets with an efficiency up to $95\%$ \cite{2015CQGra..32u5012P}. In this paper we evaluate the performance of these algorithms using glitches in real data from aLIGO. In Section 2 we provide details of the detector data. In Section 3 we give a brief overview of the three different algorithms and details of any improvement they underwent since the previous study. In Section 4 we present the results for the three algorithms on glitches from aLIGO L1 and H1 detector data. This is followed by a discussion in Section 5 of our plans for future improvements and classification during the second aLIGO run (O2) and first Virgo observation run.   

\begin{figure}[!t]
\begin{centering}
	\includegraphics[width=\textwidth]{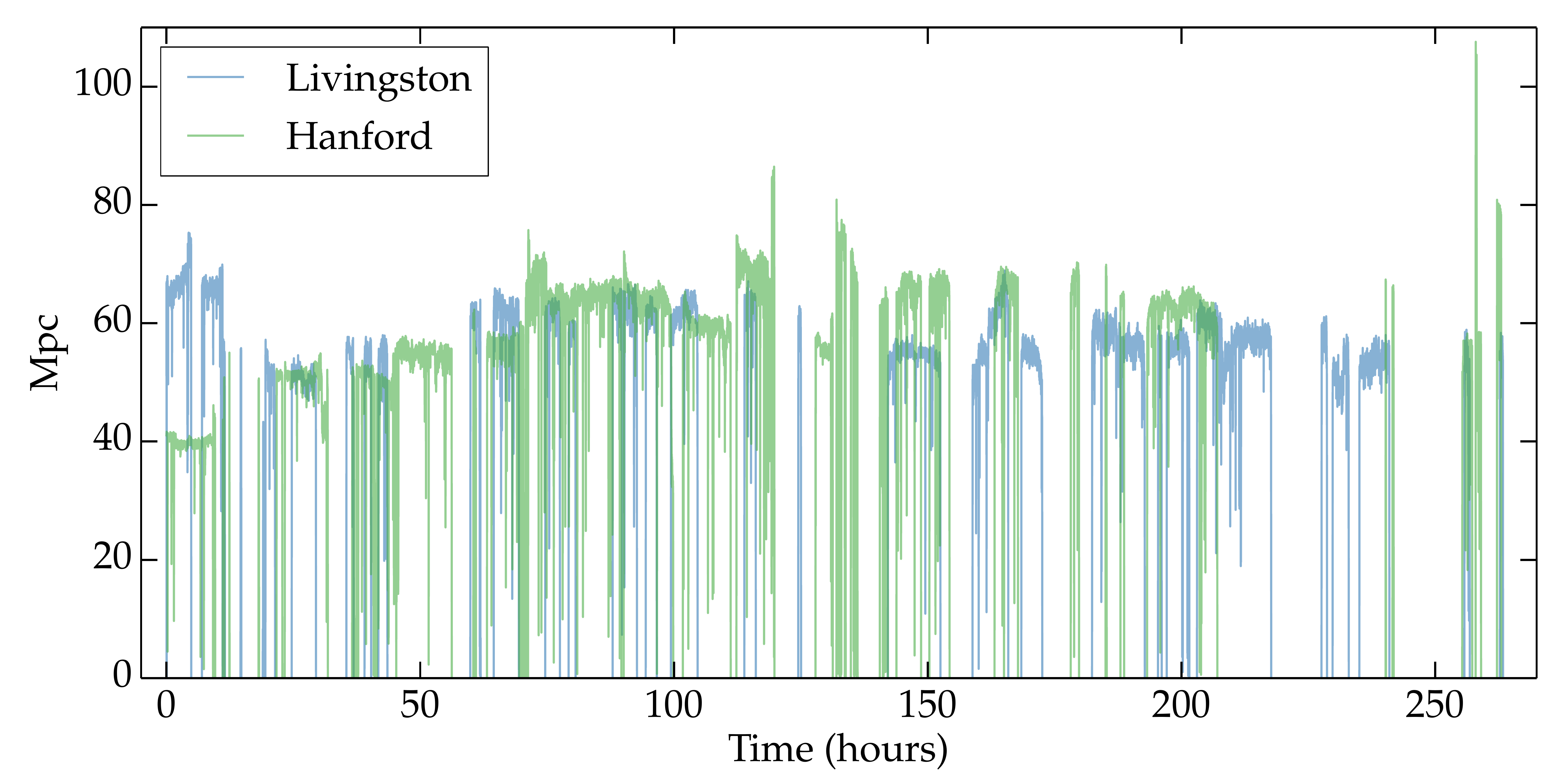}
	\label{fig:range}
\caption{The mean binary neutron star inspiral range for the two aLIGO detectors during ER7. The Hanford detector had a higher range but also a higher glitch rate. The average range was 50-60 Mpc.}
\end{centering}
\end{figure}


\section{The Data}
\label{section:dataset}

In this study we use data from the 7th aLIGO engineering run (ER7), which began on the 3rd of June 2015 and finished on the 14th of June 2015. The average binary neutron star inspiral range for both H1 and L1 detectors in data analysis mode during ER7 was $50-60$ Mpc \cite{dqshift}. The mean range for both detectors is shown in Figure \ref{fig:range}.

\subsubsection{Livingston.}

\begin{figure}[!t]
	\subfigure[]{\label{fig:scanl_a}\includegraphics[width=8.00cm]{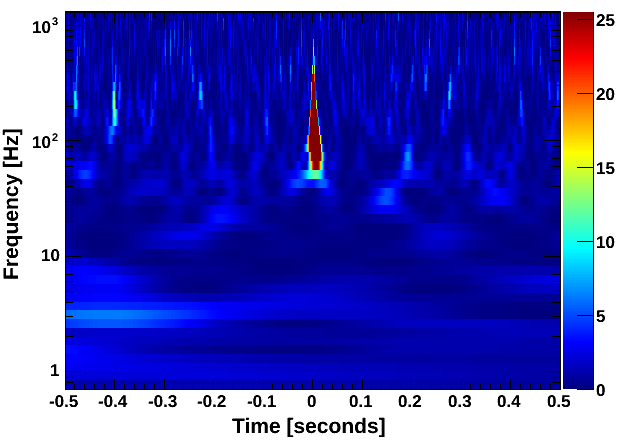}}
        \subfigure[]{\label{fig:scanl_b}\includegraphics[width=8.00cm]{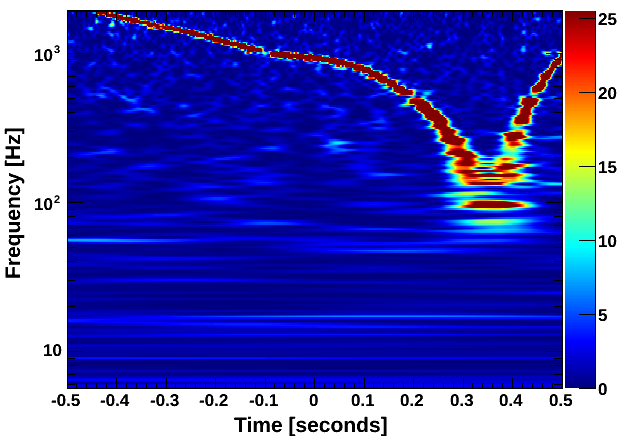}}
        \subfigure[]{\label{fig:scanl_c}\includegraphics[width=8.00cm]{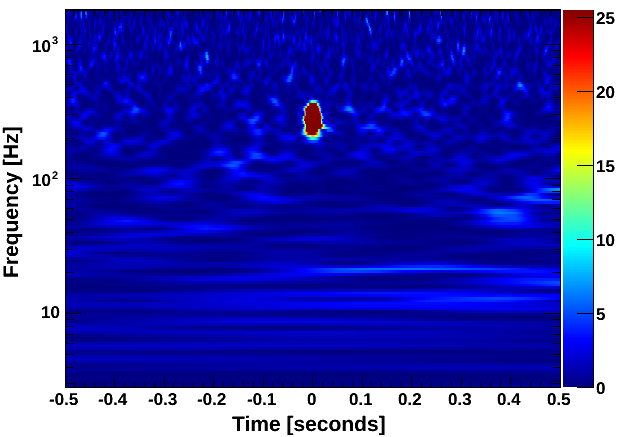}}
        \subfigure[]{\label{fig:scanl_d}\includegraphics[width=8.00cm]{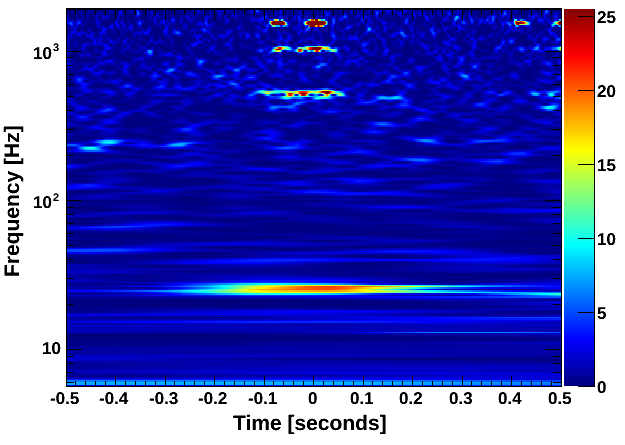}}
\caption{Spectrograms of typical transient types found in the aLIGO Livingston ER7 data. They are generated using the Omega scan tool in LigoDV-Web \cite{dvweb}, which matches the data to sine Gaussians.  (a) A transient  characterized by a tear drop shape in the spectrogram. (b) A ``whistle" glitch that often has a long duration and occurs at high frequencies. (c) A hardware injection. (d) A transient type characterized by high frequency lines and lower frequency features.}
\label{fig:scans_liv}%
\end{figure}

In the period analyzed, data from L1 consists of 48 segments where the interferometer was locked and in data analysis ready mode. These data segments vary in length from
$1$ second to $\sim\,7\,$hours. We discard any segments of data that are less than a minute in duration as a longer segment of data is required
to measure the power spectral density (PSD). The total discarded amount was $49\,$seconds of data. The total length of L1 data analysed is $\sim\,87\,$hours.

Glitches of different types are often recognised by their shape in a spectrogram such as those shown in Figure \ref{fig:scans_liv}. A description of the most common glitch types, which have occurred in aLIGO data, are described in \cite{classes}. Figure \ref{fig:scans_liv}\subref{fig:scanl_a} shows glitches characterized by a tear drop shape. Figure \ref{fig:scans_liv}\subref{fig:scanl_b} shows longer duration transients known as ``whistles", which are caused by radio frequency beats \cite{classes}. Only a small number of whistles $(\sim\,11)$ were found in the frequency and SNR range used in this study. Some other glitches in the data that are not shown in Figure \ref{fig:scans_liv} include those below $10\,$Hz and scattered light. Glitches span the entire frequency range considered in this study. Some transients may have occurred due to the increased microseism created by tropical storm ``Bill" in the Gulf of Mexico \cite{dqshift}.

A number of hardware injections were also made during ER7. An example is shown in Figure \ref{fig:scans_liv}\subref{fig:scanl_c}. Hardware injections are artificial signals simulated by inducing a motion of the optics that can be used to test which auxiliary channels are sensitive to gravitational waves \cite{2011CQGra..28w5005S, 2014PhRvD..89l2001A}.

\subsubsection{Hanford.}

\begin{figure}[!t]
	\subfigure[]{\label{fig:scanh_a}\includegraphics[width=8.00cm]{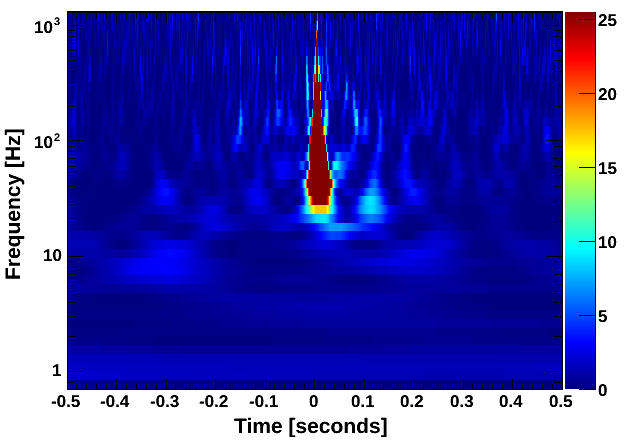}}
        \subfigure[]{\label{fig:scanh_b}\includegraphics[width=8.00cm]{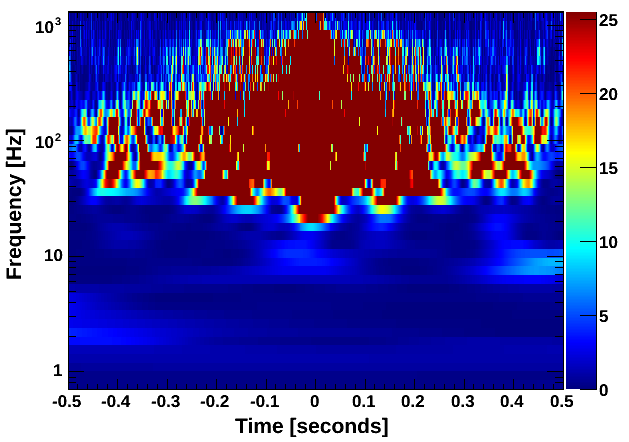}}
        \subfigure[]{\label{fig:scanh_c}\includegraphics[width=8.00cm]{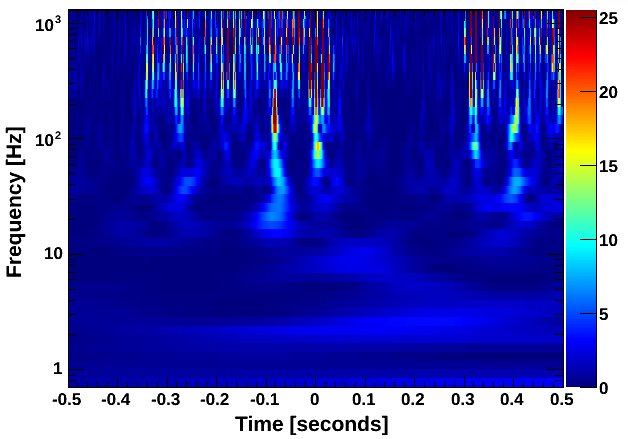}}
        \subfigure[]{\label{fig:scanh_d}\includegraphics[width=8.00cm]{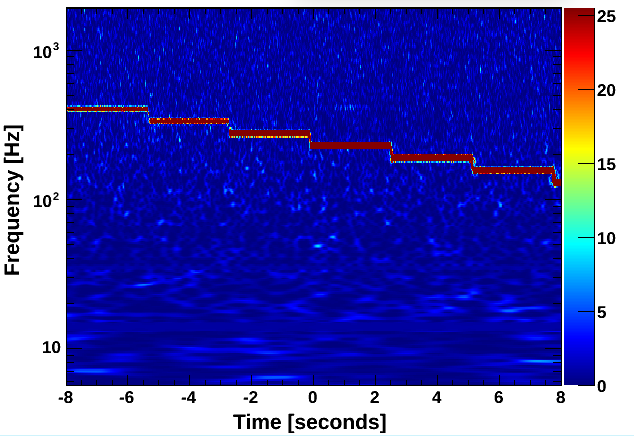}}
\caption{ Examples of some of the most common transient types found in the Hanford ER7 data (a) A tear drop glitch. (b) Transients of this type have a large SNR and duration. They created significant drops in the detectors range. (c) A high frequency transient type. (d) A longer duration line occurring at the beginning of a number of data segments. }
\label{fig:scans_hanf}
\end{figure}

In the period analyzed, data from the H1 detector consists of 50 segments where the interferometer was locked and in data analysis ready mode. The data segments vary in length from
$1\,$second to almost $14\,$hours. As with L1 we discard any segments of data that are less than a minute in duration, which was a total of $116\,$seconds of data. The total length of Hanford data analysed is $\sim\,141\,$hours.

The H1 data is highly non-stationary and contains many more transients than the aLIGO L1 data. In particular, the H1 data contains many high SNR transients that caused a significant drop in the binary neutron star inspiral range. An example is shown in Figure \ref{fig:scans_hanf}\subref{fig:scanh_b}. It was suspected that these large transients were caused by cleaning of the beam tube \cite{dqshift}. A few other examples of common transients found are shown in the other spectrograms displayed in Figure \ref{fig:scans_hanf}. As with the L1 data, H1 data also contains a number of hardware injections.


\section{Transient classifying algorithms}
\label{section:algorithms}

\begin{figure}[!t]
	\subfigure[]{\label{fig:pipe_pcat}\includegraphics[width=4.70cm]{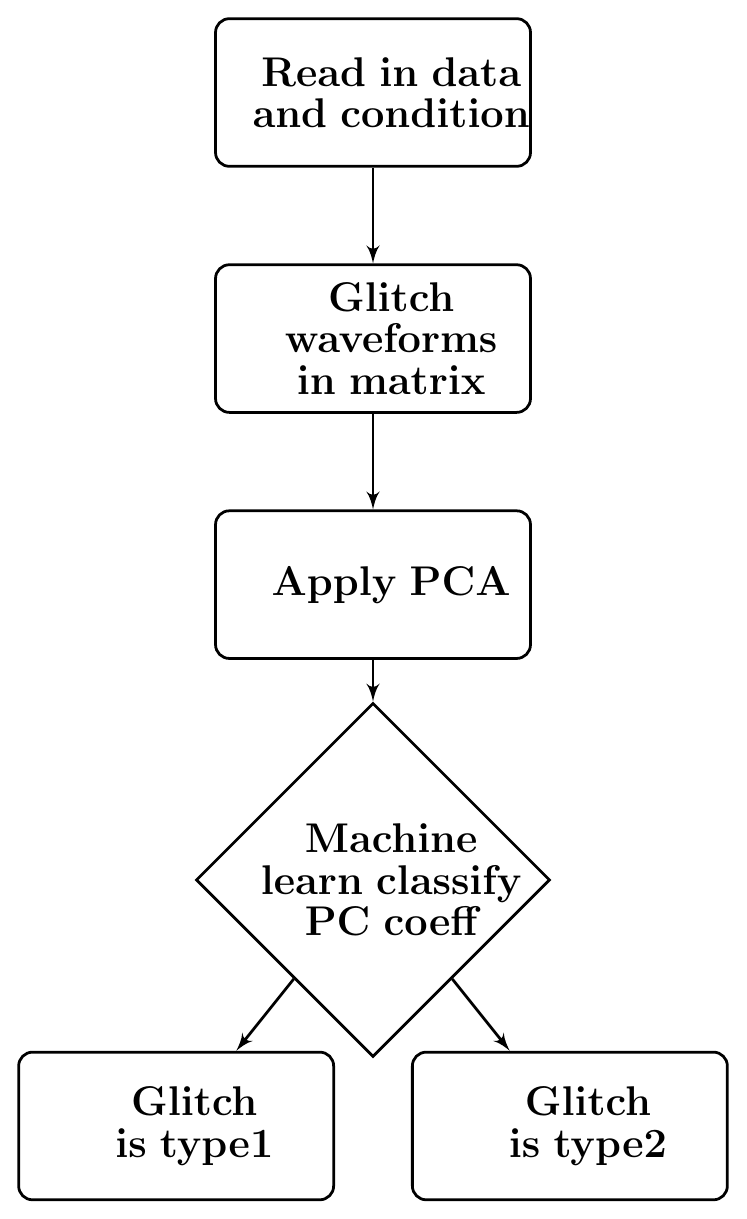}}
        \subfigure[]{\label{fig:pipe_lib}\includegraphics[width=6.00cm]{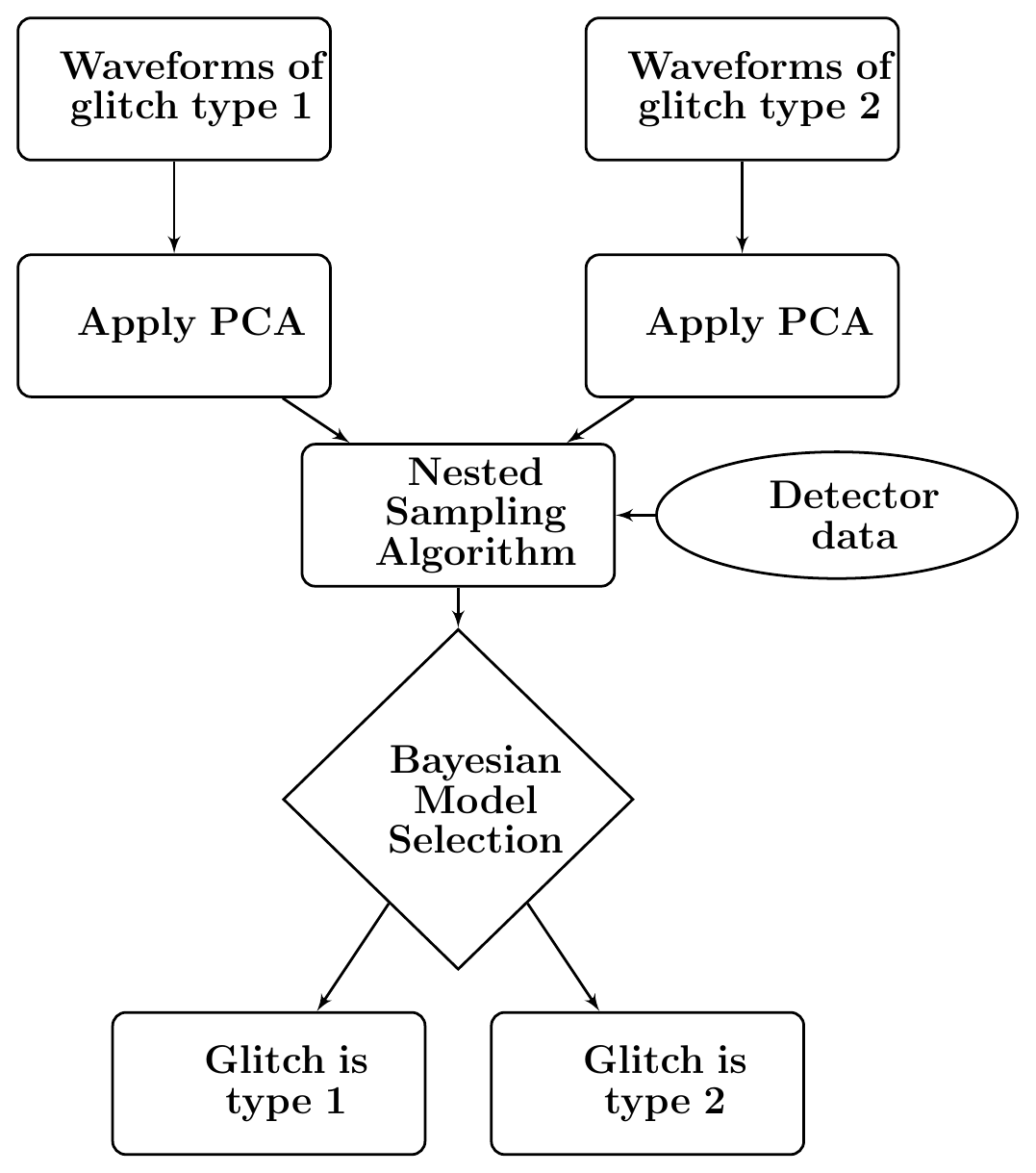}}
        \subfigure[]{\label{fig:pipe_wdf}\includegraphics[width=4.70cm]{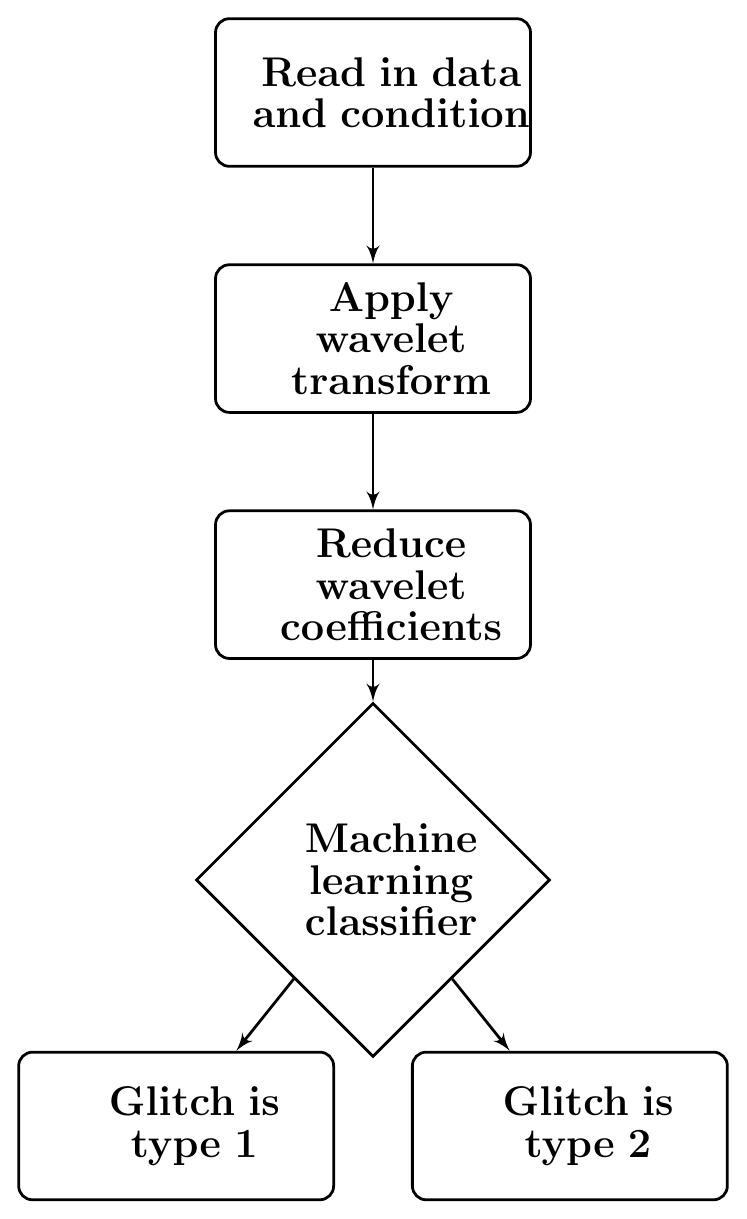}}
\caption{Classification procedures for the three different methods used in this study. (a) PCAT applies principal component analysis to all transients detected in a stretch of data and then applies a machine learning classifier to the principal component coefficients. (b) PC-LIB uses a combination of principal component analysis and Bayesian model selection to determine the glitch type. (c) WDF-ML applies a machine learning classifier to wavelet coefficients obtained by applying a wavelet transform to the transients in the data.}
\label{fig:pipeline}
\end{figure}

Three different classifying algorithms were developed for the fast classification of noise transients in the detectors. Most of the technical details have been described in \cite{2015CQGra..32u5012P}. Here we give a brief outline of the three methods and describe any changes that have been made to improve their performance and latency. Figure \ref{fig:pipeline} outlines the classification procedures for all three methods. More details are given in the following subsections.

To find transients in the data we use event trigger generators (ETGs). ETGs typically search for excess power in individual interferometers and output the time, SNR, frequency, duration and other parameters of transients found in the data. PC-LIB uses Omicron, the main ETG used by the LIGO Scientific Collaboration's (LSC) detector characterization group \cite{Omicron, 2004CQGra..21S1809C}. WDF-ML and PCAT have their own internal ETGs.

\subsection{PCAT}
PCAT uses a technique called Principal Component Analysis (PCA) that allows for dimensional reduction of large data sets \cite{PCA-clustering, 2015CQGra..32u5012P}. In the first stage of the PCAT analysis, the data are downsampled to $8192\,$Hz, whitened and high-pass filtered at $10\,$Hz. Then PCA is applied to all of the noise transients found by the ETG in all the analyzed segments of data. PCAT uses a $0.125\,$s window around each GPS time as glitches are typically of ms duration. This can lead to a loss of sensitivity to longer duration glitches.
However, this effect can be safely neglected as longer duration glitches
do not occur very often during observing runs, when the data is generally
more stable than the ER data.

A projection of the original waveforms on to the Principal Components (PCs) allows for the calculation of scale factors for each PC called PC coefficients. Noise transients of different types are separated in the PC coefficient parameter space. This allows PCAT to classify the transients by applying a Gaussian Mixture Model (GMM) machine learning classifier to the PC coefficients \cite{scikit-learn}.

\subsection{PC-LIB}
LALInference Burst (LIB) is a Bayesian parameter estimation and model selection tool, which uses a sine-Gaussian signal model to estimate parameters of gravitational-wave bursts \cite{2014arXiv1409.2435E}. It can also be combined with Omicron to be run as a search \cite{2015arXiv151105955L}. PC-LIB adapts LIB for the classification of transients by replacing the LIB sine-Gaussian signal model with a new signal model created from a linear combination of PCs calculated from the waveforms of known transient types \cite{2009CQGra..26j5005H, 2012PhRvD..86d4023L}. These known transients may have been previously classified by examining spectrograms of the transients or by one of the other methods. Thus PC-LIB can only classify transients that have occurred in the data many times before. When transients of a new type start to appear in the data new signal models must be created. 

In our previous study we created signal models using fifty transient waveforms. In this study we only use ten waveforms. This change will allow us to start classifying new transient types more quickly as they start to appear in O2 data, without any loss in sensitivity, as glitch waveforms for specific types do not have much variance in shape. Bayesian model selection can then be used to determine what population of noise transient each new glitch belongs to \cite{Sivia1996, 2014arXiv1409.7215V, 2015arXiv151105955L}. First, one second of data around the trigger time is downsampled to $8192\,$Hz and a $10\,$Hz high pass filter is applied. Nested sampling is then used to calculate Bayes factors to determine the correct transient type \cite{Sivia1996}.

\subsection{WDF-ML}

Wavelet detection filter (WDF) is an ETG that is part of the Noise Analysis Package (NAP), developed by the Virgo collaboration \cite{NAP, WDF-GRB}. It is combined with a machine learning classifier for transient classification (WDF-ML). 
 
In order to reduce the number of wavelet coefficients produced by WDF-ML, the data are downsampled before any data conditioning in the time domain to prevent border effects introduced by the Fast Fourier Transform (FFT). The downsampling is a new feature of WDF-ML that was not implemented in the version of the algorithm used in our previous study. The data are then whitened using parameters estimated at the beginning of each locked segment. After whitening, a wavelet-transform is applied, using a bank of wavelets, as described in \cite{2015CQGra..32u5012P}. We use a window of $2048$ points, with an overlap of $1968$ points, which corresponds to a duration of $0.25\,$seconds, as transients are typically of a short (ms) duration. 

The wavelet coefficients identified by the WDF-ML ETG are further cleaned using a wavelet de-noising procedure where only wavelet coefficients above the noise level are retained \cite{2015CQGra..32u5012P}. WDF-ML produces a list of wavelet coefficients, frequency, duration and SNR for each transient. The dimensions of the wavelet coefficients are then reduced by applying PCA and Spectral Embedding \cite{Ng01onspectral, belkin2003laplacian}. The transient classification is then performed by applying a machine learning GMM classifier to the reduced wavelet coefficients \cite{scikit-learn}.


\section{Classification}
\label{section:Results}

In the following sections we show the classification results obtained by PCAT, PC-LIB and WDF-ML on aLIGO H1 and L1 data. All algorithms are run with the same configurations that we expect to use during O2 to better understand our performance during the future observing runs. To determine if the glitches are classified correctly spectograms of all glitches are made and visually inspected. 

\subsection{Livingston}

To find the transients in the L1 data we look for triggers that are coincident within half a second in the outputs of all ETGs. The WDF-ML ETG was run with an SNR threshold of 10 at a sampling rate of $8192\,$Hz. Omicron was run with a lower SNR threshold of 5. We then look for transients that are coincident between both WDF-ML and Omicron, above SNR 20, and find a total of 426 coincident transients. As the PCAT ETG cannot find the lower frequency (below $10\,$Hz) triggers and some longer duration triggers we still classify transients that are coincident between Omicron and WDF, but missed by PCAT, as those triggers would still be classified when running in low latency.  

\begin{figure}[!t]
\begin{centering}
	\subfigure[]{\label{fig:wavel_a}\includegraphics[width=5.12cm]{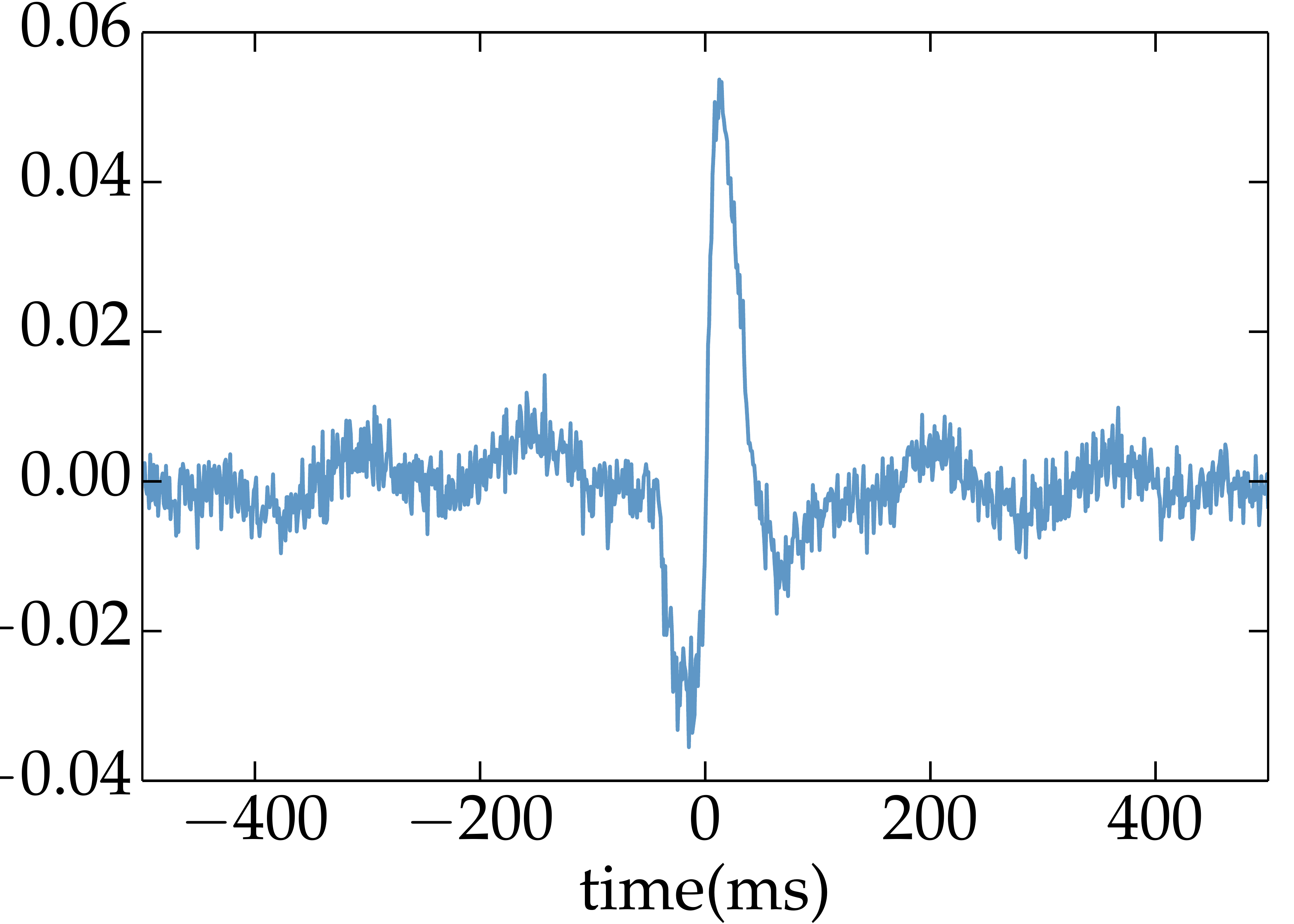}}
	\subfigure[]{\label{fig:wavel_b}\includegraphics[width=5.12cm]{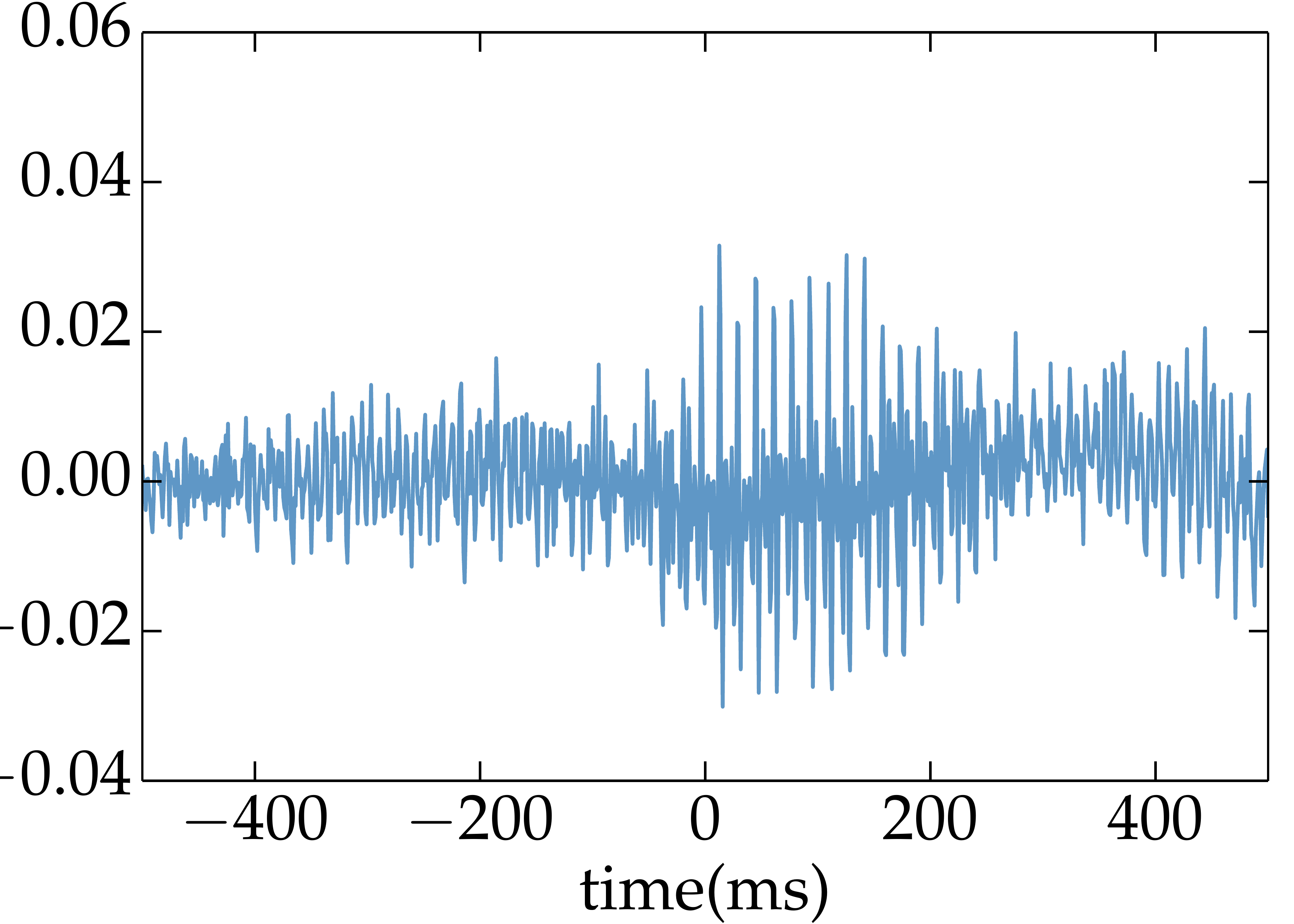}}
	\subfigure[]{\label{fig:wavel_c}\includegraphics[width=5.12cm]{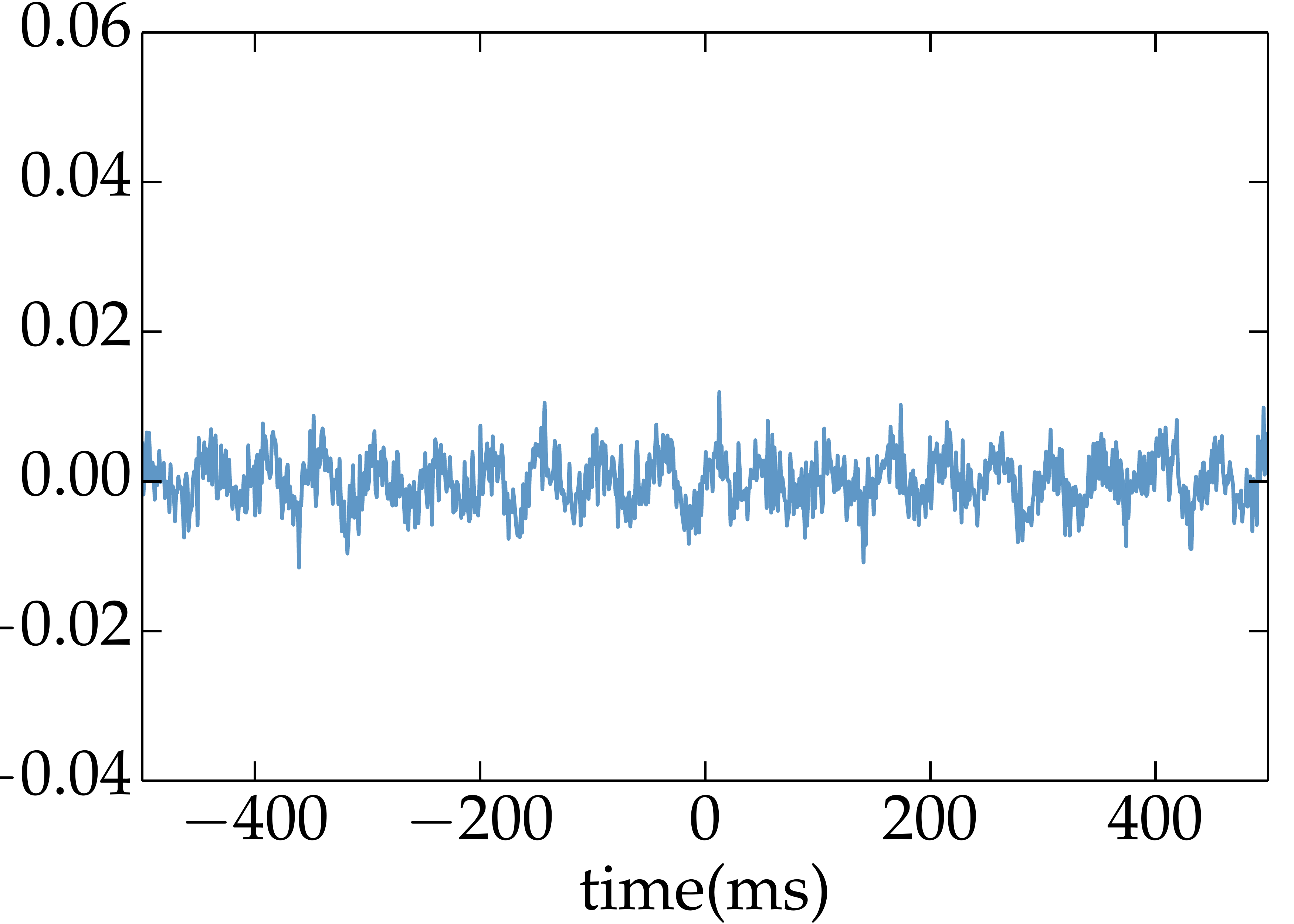}}
\caption{ The typical high pass filtered and whitened time series waveforms for three of the most common transient types found in the Livingston detector. (a) A spike which appears as a tear drop in a spectrogram. (b) The time series waveform of the glitches shown in Figure \ref{fig:scans_liv}\subref{fig:scanl_d}. (c) The time series of a whistle glitch.}
\label{fig:waveforml}
\end{centering}
\end{figure}

\subsubsection{PCAT}

applied a threshold on the SNR of the transients of 4.5 and the maximum possible number of transient types was set to 10. The ideal number of PCs can be estimated by finding the knee of the data set variance curve. This gives a total number of 20 PCs. 

PCAT classifies all the transients into 10 different classes. 90 triggers that were coincident between the Omicron and WDF-ML ETGs were missed by the PCAT ETG. Included in these missed triggers are all of the whistles, as their duration is longer than the PCAT analysis window, and 17 transients that are not visible in a spectrogram. 20 of the lower SNR hardware injections are also missed.

The data contains three main types of transients with examples shown in Figure \ref{fig:scans_liv}\subref{fig:scanl_a}, \subref{fig:scanl_b} and \subref{fig:scanl_c}. As PCAT does not detect any of the whistles shown in Figure \ref{fig:scans_liv}\subref{fig:scanl_b} the remaining glitches are classified into two main types, further split into sub-types. 

PCAT classes 1, 4 and 10 contain the transients which appear as a spike in the time series, as shown in the \ref{fig:waveforml}\subref{fig:wavel_a} and in the spectrogram in Figure 2(a). Class 4 contains only 2 transients, class 1 contains 123 transients and class 10 contains 100 transients. Class 1 and 10 contain 11 and 20 hardware injections respectively. The three sub classes are characterized by different duration of the transients. Triggers in class 1 have the lowest $(\leq 0.005\,$s) duration, class 10 have a larger $(\leq 0.01\,$s) duration, and class 4 contains two longer $(\geq 0.01\,$s) duration spikes. Two of the transients in class 10 were incorrectly classified. 

Classes 3, 5, 6, 7 and 8 contain the transients with a time series waveform shown in Figure \ref{fig:waveforml}\subref{fig:wavel_b} and a spectrogram shown in Figure 2(d). Triggers in classes 5, 7 and 8 all have SNR values between $20$ and $25$ and durations of $\sim\,0.01$ s. Class 3 contains triggers of the same transient type but with larger durations $(\leq 0.02$ s) and SNR values up to $50$. Class 6 contains only one transient, also of the same type, but with an SNR value of $57$ and a duration value of $0.005$ s.

PCAT classes 2 and 9 contain 11 and 7 glitches respectively. As these transients are not visible in a spectrogram it is not possible to determine what their type is and if they are classified correctly. Overall $95\%$ of the transients are correctly classified by PCAT.

\subsubsection{PC-LIB}

classifies all transients into four different types. To create the signal models the first 5 PCs for each glitch type are used as determined by the knee of the explained variance curve. Class 0 contains 33 transients that are not detected by PC-LIB and are thus classified into a noise class. Most of the noise class transients occur at frequencies lower than the $10\,$Hz cutoff used by PC-LIB.  

Class 1 contains 249 transients that correspond to PCAT class 4 and 5 and appear as a spike in the time series. Two of the transients in this type are mis-classified. All of the hardware injections in the data are classified in this type. Class 2 contains 131 transients which correspond to PCAT sub-classes 3, 5, 6 and 7 and have a time series waveform shown in Figure \ref{fig:waveforml}\subref{fig:wavel_b}. There are no incorrectly classified transients in this class.    

Finally, class 3 contains 13 transients. Most of the transients in this class are the whistle transients shown in Figure 2(b) and in Figure \ref{fig:waveforml}\subref{fig:wavel_c}. Three of the transients in this class are mis-classified and should be in class 2. Overall PC-LIB classifies $98\%$ of the detected transients correctly.

\begin{figure}[!t]
	\subfigure[]{\label{fig:compl_a}\includegraphics[width=8.00cm]{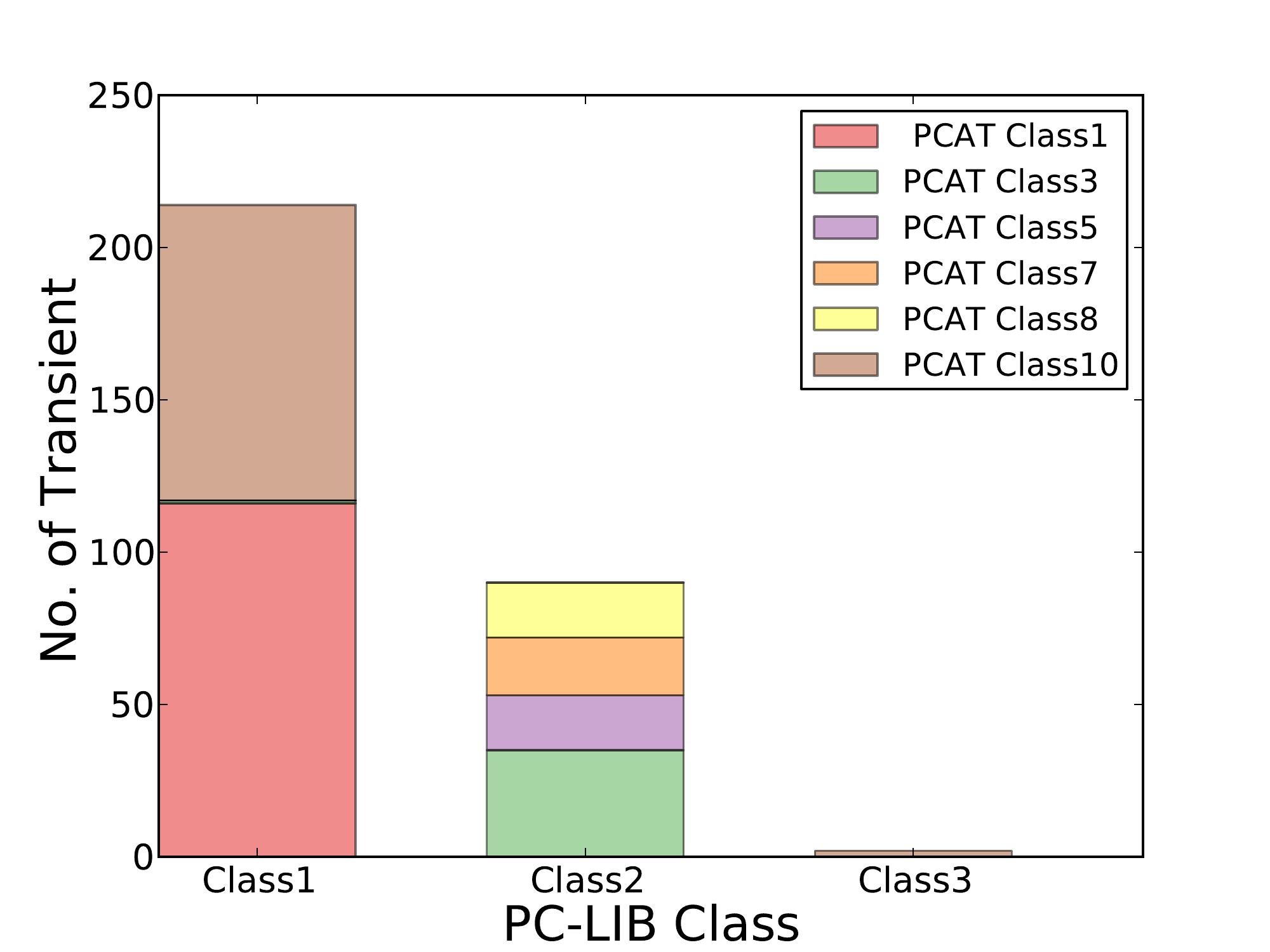}}
        \subfigure[]{\label{fig:compl_b}\includegraphics[width=8.00cm]{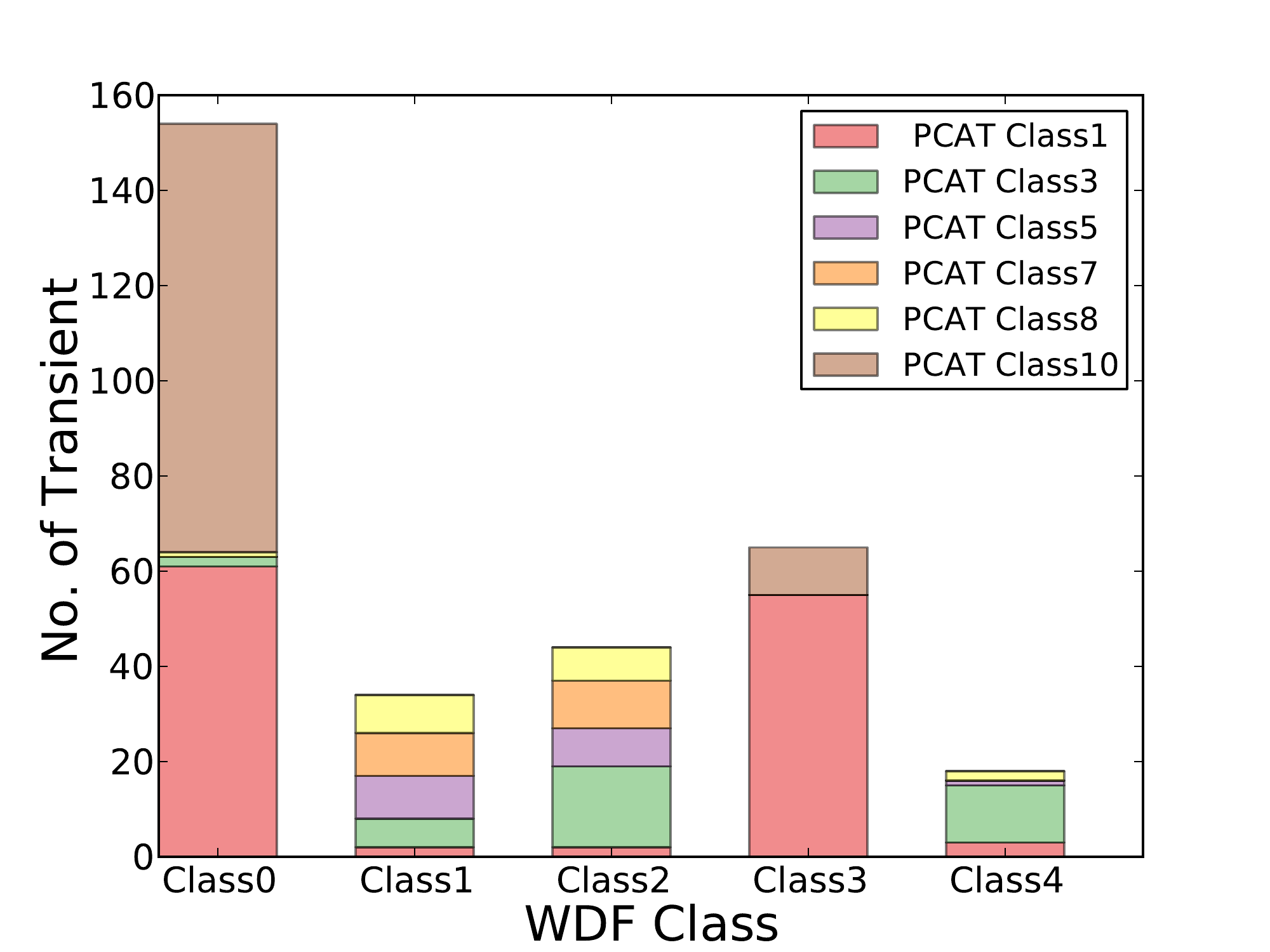}}
        \subfigure[]{\label{fig:compl_c}\includegraphics[width=8.00cm]{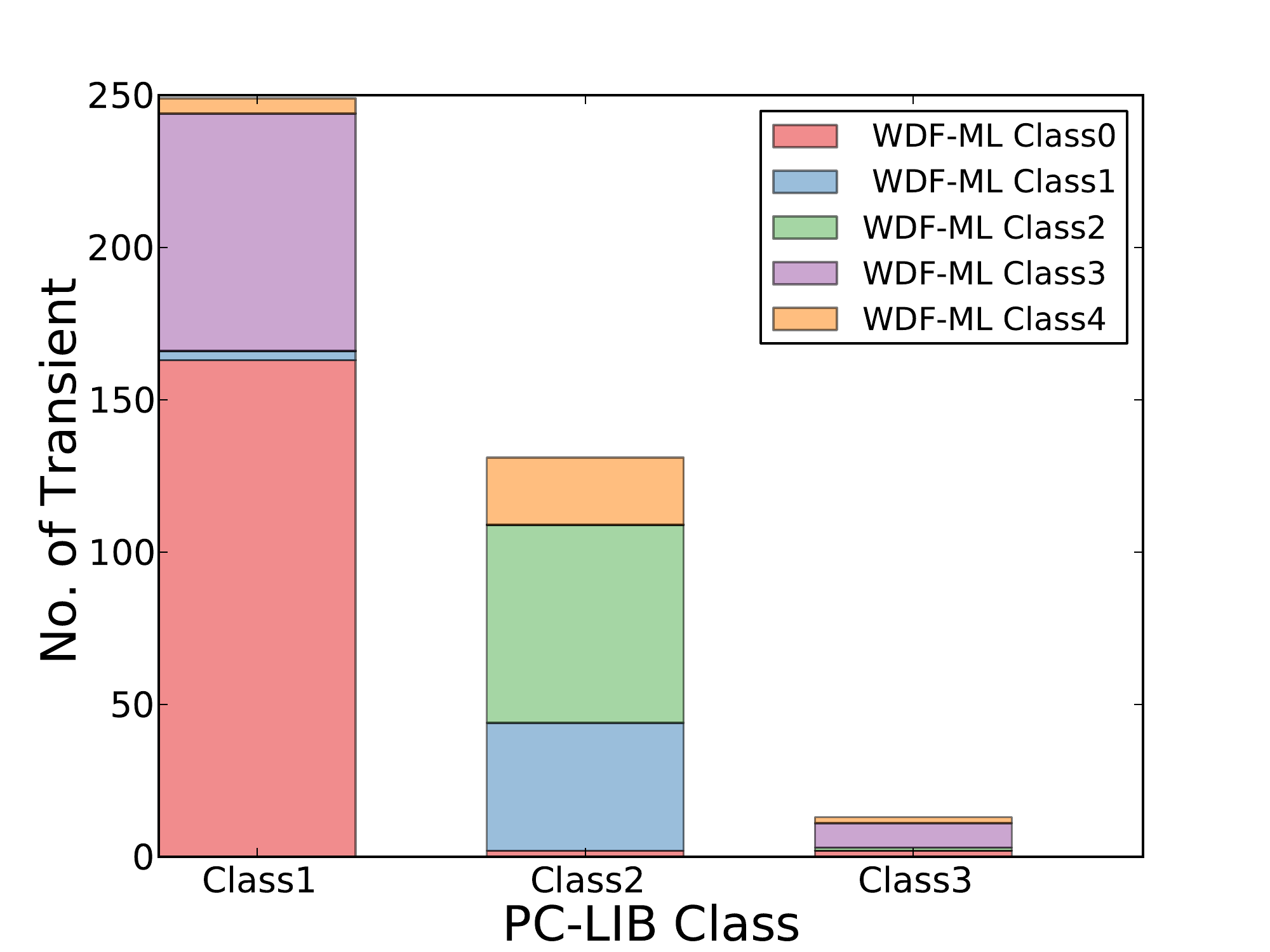}}
        \subfigure[]{\label{fig:compl_d}\includegraphics[width=7.50cm]{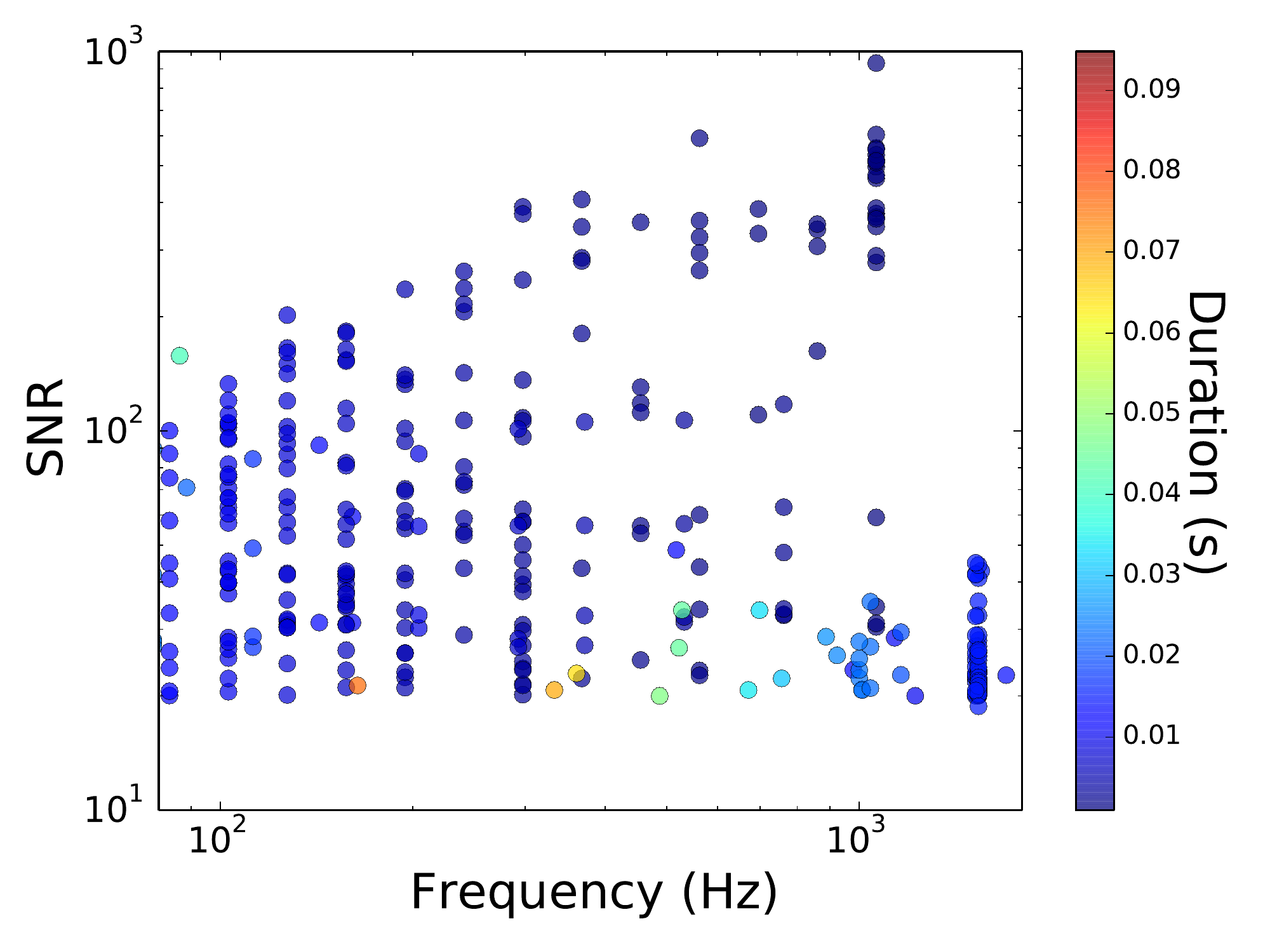}}
\caption{Classification comparisons for the three different methods on the data from LIGO Livingston. (a) Compares the classification results of PCAT and PC-LIB. PCAT class 2,4,6 and 9 are not shown as they contain less than 15 transients. (b) Compares the classification results of PCAT and WDF-ML. (c) Compares the classification results of PC-LIB and WDF-ML. (d) The SNR and frequency of all the transients classified in the data.}
\label{fig:comparel}
\end{figure}  

\subsubsection{WDF-ML}

classifies all transients into five different classes. The 5 classes consist of two different types of transients as WDF-ML cannot accurately classify the longer duration whistles due to the short analysis window. Sub-classes are determined by the wavelet family of the transients rather than split by duration or SNR as for PCAT. 

WDF-ML class 0 contains 195 transients and class 3 contains 86 transients that appear as a spike in the time series. These correspond to PCAT types 1, 4 and 10 and PC-LIB's class 1. The two sub-classes contain 29 hardware injections. They also contain 8 of the whistle glitches, as WDF-ML cannot accurately classify longer duration transients. Four of the class 0 and one of the class 3 transients are incorrectly classified. 

The second main transient type found by WDF-ML corresponds to PC-LIB's class 2 and contains transients characterized by the typical spectrogram shown in Figure 2(d). The transients were split into three sub-classes, namely class 1 that contains 46 transients, class 2 that contains 70 transients and class 4 that contains 29 transients. Class 1 contains three incorrectly classified transients and class 3 contains two of the whistles glitches. Class 4 contains 4 hardware injections that are mis-classified. Overall WDF-ML classifies $95\%$ of the L1 transients correctly.

\subsubsection{Comparison.}

Figure 6 shows a comparison of the classifications made by all three methods. All methods are able to classify transients with a high level of accuracy in real non-stationary data. WDF-ML performs better at classifying very low frequency transients as it does not need to use a lower frequency cutoff. The Omicron SNR, duration and frequency of all 426 transients are shown in Figure 6(d). The constant lines are due to the Omicron's method for measuring frequency \cite{Omicron}.

Only PC-LIB is able to separate the whistle transients into a separate class due to the longer $1\,$s time window used by this method. The efficiency in classifying these transients for the other algorithms could be improved by using a longer time window. However, this could lead to multiple shorter duration transients occurring in the same time window. As PC-LIB looks for specific known transient types it could be used to add labels to the classifications of the other methods so that it will make it easier to find out which class corresponds to which transient type, defined in \cite{classes}, and which classes are new types that have not occurred previously. As WDF-ML and PCAT can classify new transient types as soon as they appear in the data they can be used to provide waveforms for PC-LIB to use to create new signal models.

\subsection{Hanford}

As for the L1 data transients coincident within $0.5\,$s between all ETGs are classified. A higher SNR threshold of 30 is used for H1 as the data contains many more transients than the L1 data and is more non-stationary. A total of 1865 coincident transients are classified in H1.

\begin{figure}[!t]
\begin{centering}
	\subfigure[]{\label{fig:waveh_a}\includegraphics[width=5.12cm]{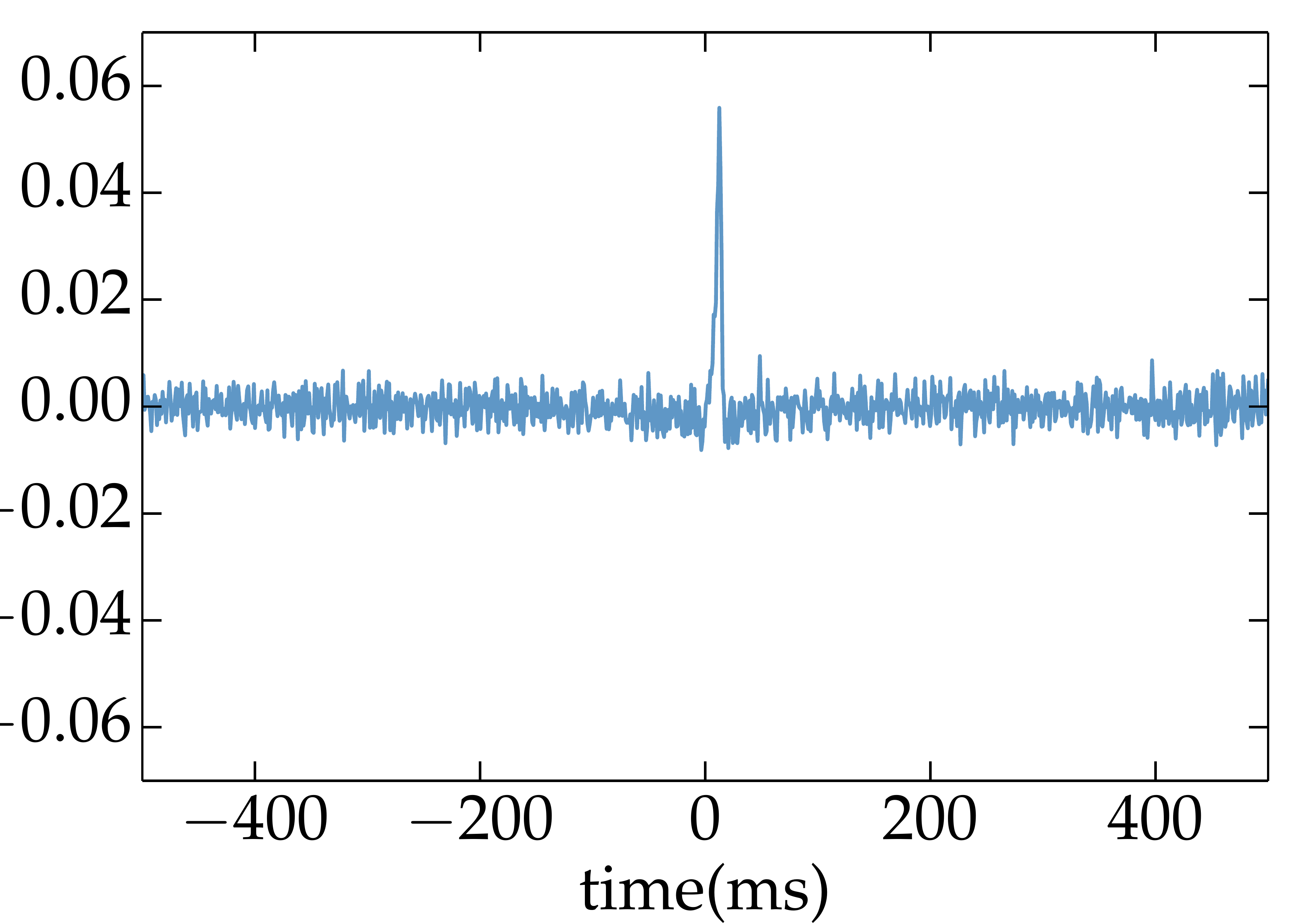}}
	\subfigure[]{\label{fig:waveh_b}\includegraphics[width=5.12cm]{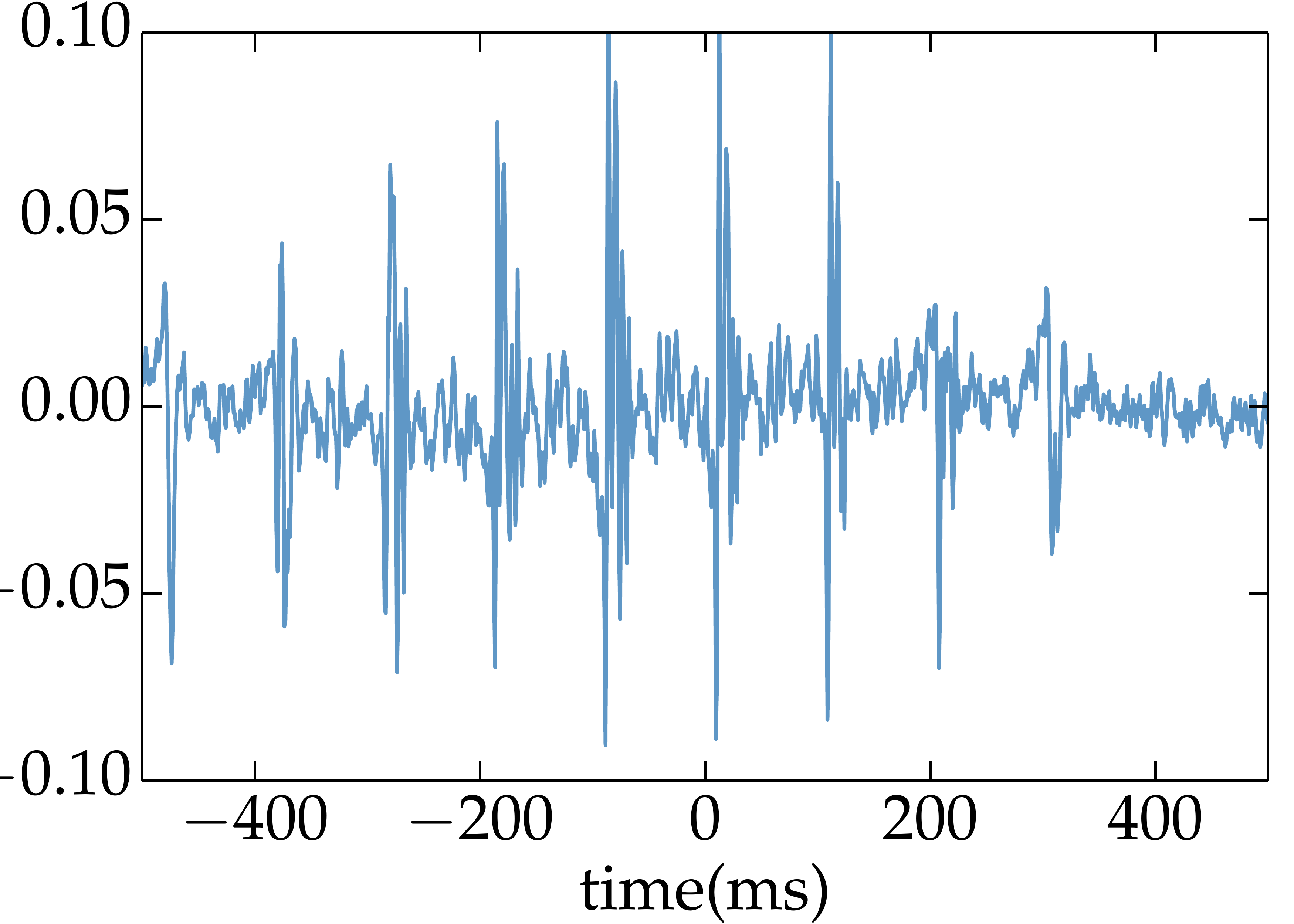}}
	\subfigure[]{\label{fig:waveh_c}\includegraphics[width=5.12cm]{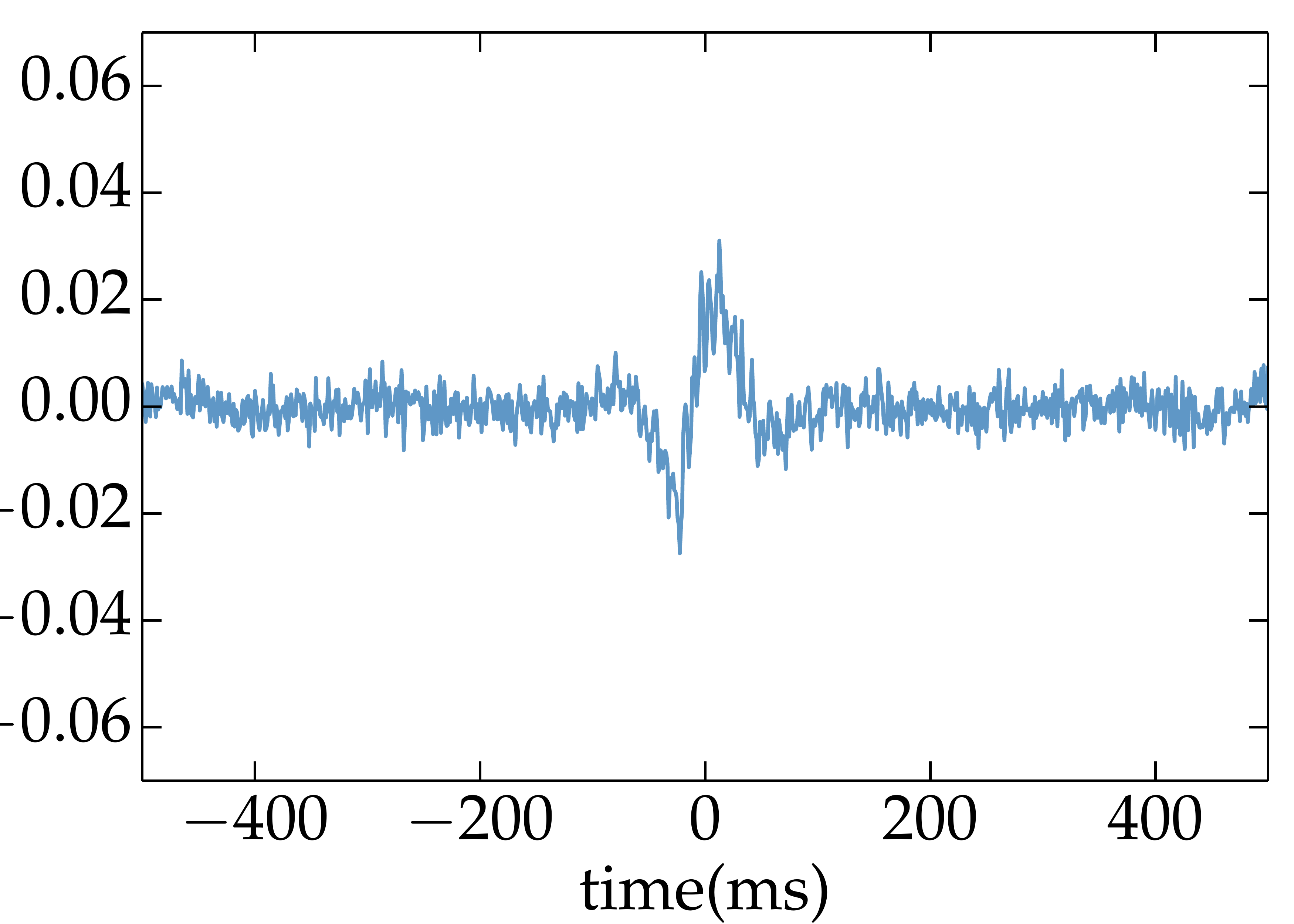}}
\caption{ The typical whitened and high pass filtered time series waveforms for three of the most common transient types found in the Hanford detector. (a) A spike in the time series that appears as a tear drop in a spectrogram. (b) The time series of second most common glitch found in the data, as shown in Figure 3(c). (c) The time series waveform of a hardware injection. }
\label{fig:waveformh}
\end{centering}
\end{figure}

\subsubsection{PCAT}

uses 20 PCs to classify the transients into 7 different types. 120 of the transients coincident between WDF-ML and Omicron ETGs are not detected by the PCAT ETG. They are transients below $10\,$Hz or triggers from the long duration lines, shown in Figure 3(d), which are not really glitches. The detected transients are split into 7 different classes. 

The data contains two main types of transients. The first type is characterised by a typical spectrogram shown in Figure 3(a) and a time series waveform shown in Figure \ref{fig:waveformh}\subref{fig:waveh_a}. PCAT splits this type into 6 different sub-classes with 267, 603, 648 44, 1, and 64 transients respectively. Class 1 has 9 mis-classified transients. Classes 2, 3 and 5 all have one mis-classified transient. Class 2, 3 and 6 contain lower duration $(\sim 0.005$ s), but with different Q and frequency ranges, where Q is defined as $Q={\rm duration}\times 2\pi \times {\rm frequency}$. Class 1, 5 and 16 contain relatively longer duration waveforms $(\sim 0.01$ s) which also have different Q and frequency ranges. 

The second type of transient has a typical time-frequency morphology shown in Figure 3(c) and time series waveforms shown in Figure \ref{fig:waveformh}\subref{fig:waveh_b}. This type is found in PCAT class 4 that contains 117 transients that are all classified correctly. Overall PCAT classifies $99\%$ of the detected H1 transients correctly. 

\subsubsection{PC-LIB}

As with the L1 data we use 5 PCs to create signal models for the H1 transients. PC-LIB splits the transients into two different classes. A noise class contains the 6 transients shown in Figure 3(d) as they cannot be detected. 

Class 1 contains 1651 transients that correspond to a spike in the time series as in PCAT sub-classes 1, 2, 3, 5, 6 and 7. This class also contains 13 hardware injections. 23 transients are mis-classified and should be in class 2.    

Class 2 contains 207 transients, which have a typical spectrogram shown in Figure 3(c), and correspond to PCAT class 4. This class also includes 4 hardware injections that are more similar to a sine-Gaussian in shape than those classified into class 1. This class includes 61 transients that are mis-classified and should be in class 1. Overall PC-LIB classifies $95\%$ of the detected H1 transients correctly.

\subsubsection{WDF-ML}

splits the H1 data into three different classes. 
Class 1 contains 1358 transients, which appear as a spike in the time series, and correspond to PC-LIB class 1 and the 6 PCAT sub-classes. This class contains all hardware injections and all very low frequency transients that can not be detected by PCAT and PC-LIB. 10 of the transients in this class are mis-classified. WDF-ML class 2 contains 145 transients that are characterized by spikes in the time series, but have longer durations and lower SNR values than the transients in WDF-ML class 1.  

WDF-ML class 0 contains 326 transients corresponding to PCAT class 4 and PC-LIB class 2. This class also contains 122 mis-classified transients. As before, this is because all of the transients in the class have a duration $(\sim1$ s) which is much longer than the time window used in the WDF-ML analysis. Overall WDF-ML classifies $\sim 92\%$ of the H1 transients correctly.

\subsubsection{Comparison.}

\begin{figure}[!t]
	\subfigure[]{\label{fig:comph_a}\includegraphics[width=8.00cm]{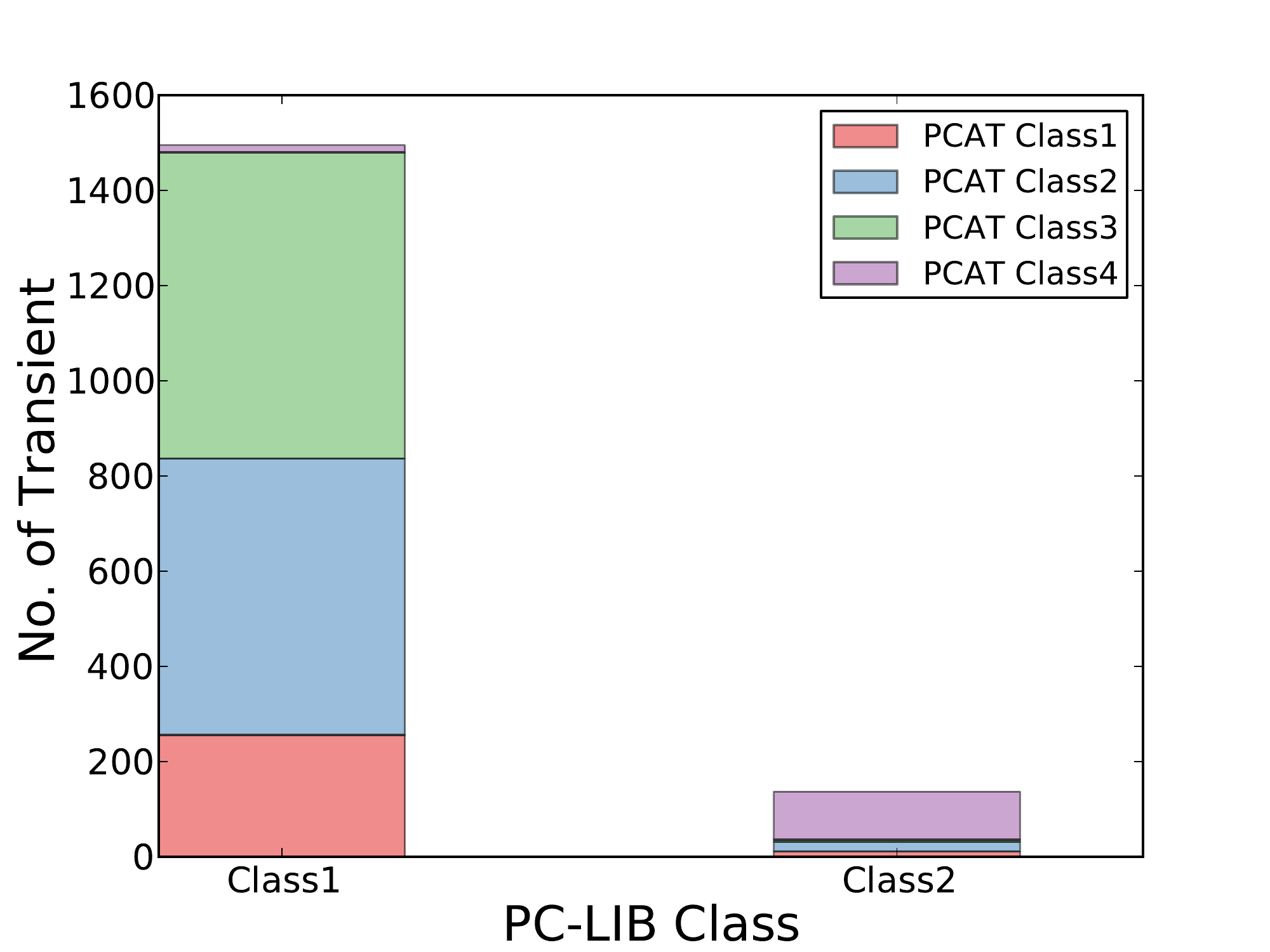}}
        \subfigure[]{\label{fig:comph_b}\includegraphics[width=8.00cm]{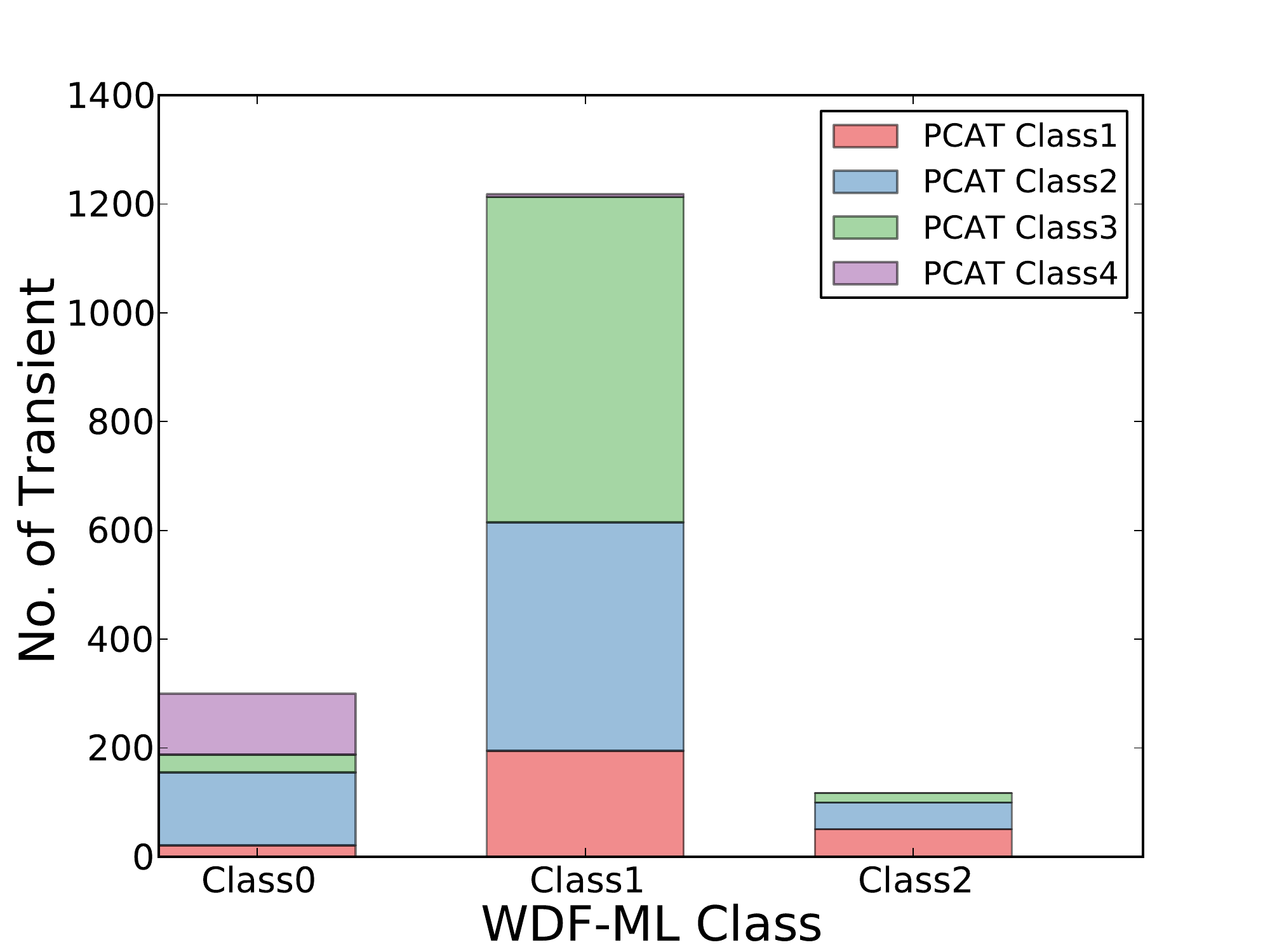}}
        \subfigure[]{\label{fig:comph_c}\includegraphics[width=8.00cm]{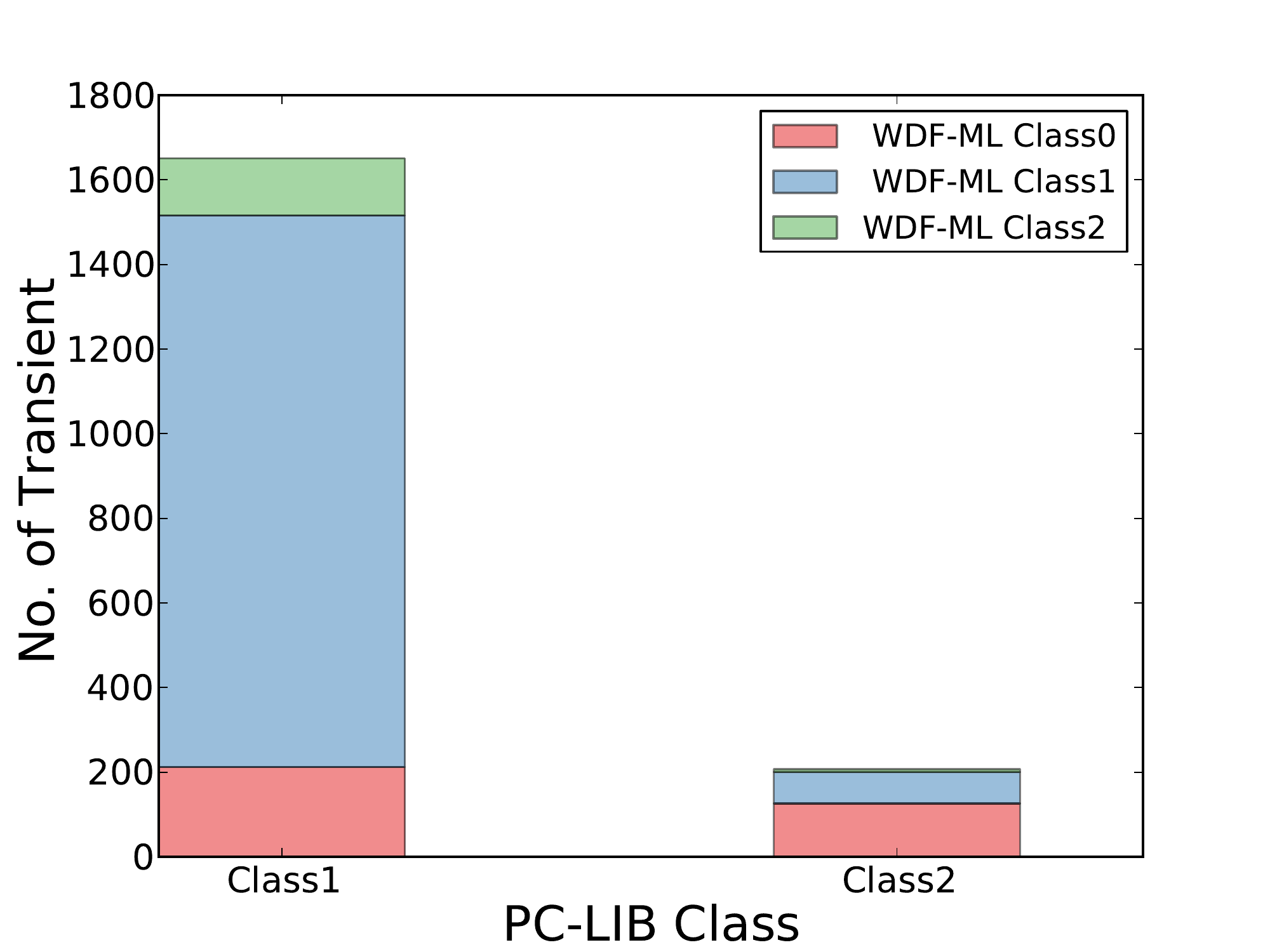}}
        \subfigure[]{\label{fig:comph_d}\includegraphics[width=7.50cm]{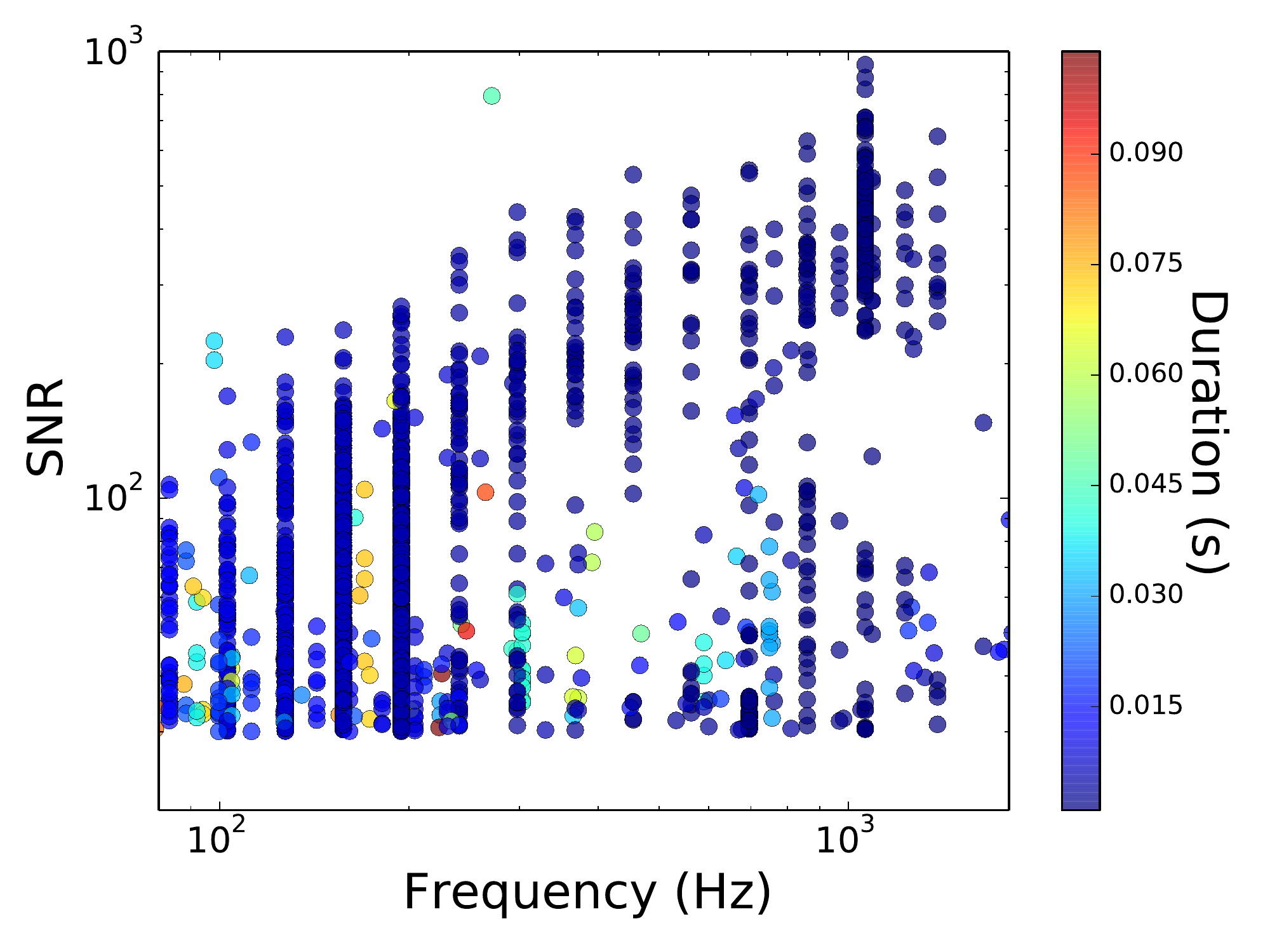}}
\caption{Classification comparisons for the three different methods for aLIGO Hanford data. (a) PC-LIB splits the transients into two classes. PCAT can split different types into sub-classes. (b) PCAT and WDF-ML comparison. WDF-ML has difficulty with transients which have a larger duration than their analysis window. (c) Comparison of PC-LIB and WDF-ML classifications. (d) The Omicron SNR, duration and frequency of all the transients classified in the data. The discreteness in frequency is due to Omicron.}
\label{fig:compareh}
\end{figure}

The results obtained by all three methods for the H1 data are compared in Figure 8. The Omicron SNR, duration and frequency of the transients is shown in Figure 8(d). As WDF-ML uses a small time window of $0.25\,$s the efficiency of the classification is reduced when the data are highly non-stationary and contain many long $(\sim1\,$s) duration transients. Even with 137 mis-classified transients the overall accuracy of the WDF-ML H1 results is $\sim 92\%$. WDF-ML estimates the PSD at the beginning of each locked segment. This may introduce errors towards the end of the segment if the data is highly non-stationary. Machine learning methods perform better when the data set analysed is large. Therefore, the larger number of glitches in H1 may have improved the classification efficiency.


\section{Discussion}
\label{section:discussion}

Non-Gaussian noise in the aLIGO and Virgo detectors can potentially mimic a gravitational-wave signal, reduce the duty cycle of the instruments and decrease the sensitivity of the detectors. Classification of different noise transient signals may help identify their origins and lead to a reduction in their number. We have developed three methods for noise classification and have previously demonstrated their performance on simulated transients in simulated Gaussian aLIGO noise \cite{2015CQGra..32u5012P}. However, as real noise from the advanced detectors is non-stationary and non-Gaussian, a better understanding of how our methods will perform during the upcoming observation runs of the advanced detectors is required. Although the detectors are typically more stable during observing runs than during ER7, we expect the types of glitches investigated in this work to be representative of the glitch classes in the upcoming observing runs.

In the ER7 data from the L1 detector PCAT missed 90 transients and classified $95\%$ of the remaining transients correctly. PC-LIB missed 33 transients and classified $98\%$ of the remaining transients correctly. WDF-ML classified all transients and $95\%$ of them were correct. In the H1 data PCAT missed 120 transients and classified $99\%$ of the remaining transients correctly. PC-LIB missed 6 transients and classified $95\%$ of the remaining transients correctly. WDF-ML classified all transients and $92\%$ of them were correct. We conclude that our methods have a high efficiency in real non-stationary and non-Gaussian detector noise.  

The efficiency of the WDF-ML algorithm is reduced for the Hanford glitches because the duration of the transients becomes much larger than the analysis window, which reduces the efficiency of the overall classification. This could be prevented  by applying a high duration cutoff to the transients found by the ETG before classification. Most high duration and SNR transients are removed by data quality vetoes. Conversely, short duration transients will be more important as they have a higher impact on the gravitational-wave search backgrounds. Since they are rarely removed by vetoes their accurate classification is crucial to improve gravitational-wave searches as an accurate categorization will allow us to search for couplings within the detector \cite{2016arXiv160203843T, 2016arXiv160203844T}. 
In the future the whitening performed by WDF-ML will be improved by using a technique known as adaptive whitening \cite{adapwhite:04}.

Because of the different strengths and weaknesses of the different methods having multiple classifiers is a winning strategy. WDF-ML can classify lower frequency transients than the other two methods. PC-LIB is better able to classify longer duration transients due to its longer analysis window. PCAT can classify new types of transients as soon as they appear in the data and thus provide transient waveforms for PC-LIB's signal models.   

Further improvements could also be made by using a training set of pre-classified waveforms or exploring the use of dictionary learning algorithms for glitch classification \cite{dictlearn}. The aLIGO gravity spy project aims to build these data sets through a citizen science program \cite{2016AAS...22810902Z, Simpson:2014:ZOW:2567948.2579215}.

\ack

We thank Salvatore Vitale, Reed Essick and the Burst and DetChar groups of 
the LIGO Scientific Collaboration for helpful discussions of this work. 
DT and MC are partially supported by the National Science Foundation
through award PHY-1404139.
ISH and JP gratefully acknowledge the support of the
UK Science and Technology Facilities Council (STFC) grant numbers ST/L000946/1 and ST/L000946/1. 
JP, ISH and EC also gratefully acknowledge the support of the Scottish Universities Physics Alliance (SUPA).
ATF and JAF gratefully acknowledge the support of the Spanish MINECO (grant AYA2013-40979-P) and the Generalitat Valenciana (PROMETEOII-2014-069). This paper has been assigned LIGO document
number LIGO-P1600263.


\section*{References}
\bibliographystyle{iopart-num}
\bibliography{bibfile}

\end{document}